\documentclass[12pt]{article}
\usepackage[english]{babel}
\usepackage{amssymb}
\usepackage{amsmath}
\usepackage{indentfirst}
\usepackage{graphicx}
\usepackage{graphics}
\usepackage{epsfig}
\usepackage{subfigure}
\usepackage{multirow}
\usepackage{color}

\newcommand{\mathsym}[1]{{}}

\def\e{{\mathrm{e}}}

\def\d{{\rm d}}
\usepackage{amsfonts}
\DeclareMathAlphabet\pazobb{\encodingdefault}{fplmbb}{m}{n}%
\renewcommand{\mathbb}{\pazobb}

\def\Naturals{\hbox{\rm I\kern-.17em N}}

\def\integers{\hbox{\rm Z\kern-.3em Z}}

\def\Racionais{\hbox{\rm Q\kern.24em
            \vrule width .05em height 1.4ex depth-.05ex\kern-.26em}}
\def\Reals{\hbox{\rm I\kern-.18em R}}

\def\Reals{\hbox{\rm I\kern-.18em R}}

\def\Complexes{\hbox{\rm C\kern-.43em
       \vrule depth 0ex height 1.4ex width .05em\kern.41em}}

\def\bfp{\boldsymbol{p}}
\def\bfr{\boldsymbol{r}}
\def\bfA{\boldsymbol{A}}
\def\bfalpha{\boldsymbol{\alpha}}
\def\bfsigma{\boldsymbol{\sigma}}
\def\bfnabla{\boldsymbol{\nabla}}
\def\Erf{{\rm Erf}}

\newcommand{\beq}{\begin{equation}}
\newcommand{\eeq}{\end{equation}}

\begin{document}
\title{A quark-meson coupling model based on  Bogoliubov's model of the nucleon}
\author{
Henrik Bohr\\
{\it \small Department of Physics, B.307, Danish Technical
University,}\\{ \it \small DK-2800 Lyngby, Denmark}\\Steven A.\ Moszkowski\\
{\sl \small UCLA, Los Angeles, CA 90095, USA} \\Prafulla K.
Panda\\{\it\small Departmenet of Physics, C.V. Raman College of
Engineering,}\\{ \it \small  Vidya Nagar, Bhubaneswar-752054, India}
\\Constan\c{c}a Provid\^encia,
Jo\~ao da Provid\^encia\\
{\it \small CFC, Departamento de F\'\i sica, Universidade de Coimbra,}\\
{\it \small  P-3004-516 Coimbra, Portugal} } \maketitle
\begin{abstract}
The quark-meson coupling model due to Guichon is  formulated on the
basis of the independent quark model of the nucleon proposed by
Bogoliubov and is applied to a phenomenological description of
symmetric nuclear matter. The model predicts, at saturation density,
the compressibility $K=249$ MeV and the quark effective mass $m_q^*=
249.1$ MeV, {the effective nucleon mass  being $M^*=747.3$ MeV}. The
predicted nucleon mass radius is $r=0.93$ fm.
\end{abstract}
\section{Introduction} About almost half
a century ago,  Bogoliubov proposed an interesting model of baryons
\cite{bogolubov}, which assumes that they are composed of quarks
bound by a linearly raising potential,  as suggested by gauge
theories.
With the help of a single phenomenological parameter, the string
tension $\kappa$, this model is able to qualitatively account for
the (dynamically generated) mass  of the nucleon, for the
corresponding magnetic moment, and for the mass-radius. The
quark-meson coupling model due to Guichon \cite{guichon}
incorporates successfully the quark degrees of freedom into a
many-body effective Hamiltonian, inspired on QCD. The aim of the
present note is to obtain a phenomenological description of hadronic
matter in the framework of a combination of both models.

The quark potential has been derived by  Baker et al \cite{baker}
using a dual-superconductor picture of QCD. The distribution of
gluon fields has been investigated on the lattice by Bissey et al.
\cite{bissey}, who have shown
that the potential increases linearly with the length of the string,
and that the Y shape configurations of the gluon flux-tube
distribution
is more favorable than the L or T configurations. In the Y shape the
strings join at some point localized inside the triangle defined by
the quarks. In the L shape, one of the quarks is at the point where
the strings join. In the T shape,  the point where the strings join
is on the line segment defined by two quarks. There are similarities
and differences with Bogoliubov's model. Since this model is an
independent quark model, in it all shapes, {L,T Y,
defined by the positions of the quarks,
have the same energy, provided the sum of the distances to the
origin is the same, this being the difference. However, the
potential energy increases linearly with the length of the string,
or with the distance to the origin, this being the similarity}.
Moreover, in Bogoliubov's model, the phenomenological string tension
turns out to be about 1/4 of the string tension in the lattice
theory.

\section{Bogoliubov's independent  quark model of the nucleon} According to Bogoliubov's
proposal, the nucleon, regarded as a system of three independent
valence quarks is described by the Hamiltonian \cite{bogolubov}
\begin{equation}\label{Ha}H=\sum_{j=1}^3\bfalpha_j\cdot{\bfp}_j
+\kappa\sum_{j=1}^3\beta_j|{\bfr}_j |,\end{equation}where the
components of $\bfalpha_j$ and $\beta_j$ denote the Dirac matrices
related to the quark $j$ and $\kappa$ is the string tension. For
simplicity, a   Coulomb term $1/|\bfr_j|$, which is included in the
so called quarkonium Cornel potential, has been omitted. The model
is admittedly incomplete since it does not accommodate the hyperfine
structure and so is unable to describe the nucleon -- $\Delta$ mass
splitting. However, such a refinement is beyond the scope of the
present note.
In the presence of a static magnetic field the Hamiltonian becomes
\begin{equation}H=\sum_{j=1}^3\bfalpha_j\cdot({\bfp}_j-q_j\bfA(\bfr_j))+\kappa\sum_{j=1}^3\beta_j
\left|{\bfr}_j
\right|,\label{HA}
\end{equation}
where $q_j$ is the charge of the quark $j$ and $\bfA$ is the
potential vector.

\subsection{The Dirac Hamiltonian and its square}
The square of the Dirac Hamiltonian
$h=\bfalpha\cdot\bfp+\beta|\bfr|$ reads
\begin{equation}\label{h^2-N}h^2=\bfp^2+\kappa^2\bfr^2-i\beta\frac{\bfalpha\cdot\bfr}{|\bfr|}\kappa.\end{equation}
It is convenient to introduce the operator
$${\tilde h}^2=(\bfp^2+\kappa \bfr^2)\left[\begin{matrix}1&0\\0&1\end{matrix}\right]+\left[\begin{matrix}\bfsigma_r&0\\0&\bfsigma_r\end{matrix}\right] \kappa,$$
which is related to $h^2$ by a unitary transformation. The operators
$h^2$ and ${\tilde h}^2$ have the same eigenvalues. Indeed, we may
write
$$h^2=\left[\begin{matrix}\bfp^2+\kappa^2\bfr^2&-i\bfsigma_r\kappa\\i\bfsigma_r\kappa&\bfp^2+\kappa^2\bfr^2\end{matrix}\right].$$So,
if $\left[\begin{matrix}\chi\\\chi\end{matrix}\right]$ is an
eigenvector of ${\tilde h}^2$, then $\left[\begin{matrix}\chi\\i\chi
\end{matrix}\right]$ is an eigenvector of $h^2$.

As an approximation, we may identify the mass of the quark with the
square root of the lowest eigenvalue of $\bfp^2+\kappa^2\bfr^2$,
that is, with $\sqrt{3\kappa}$. However, we must correct for the
center of mass motion. If this is done, as explained later, the
corrected mass is $\sqrt{{5\kappa/ 2}} $.

The relevant eigenvectors of
$\bfp^2+\kappa^2\bfr^2$ read
$$\Psi_{(1;{1\over2},0,{1\over2})}=\left({\kappa\over\pi}\right)^{{3\over4}}\e^{-{1\over2}\kappa r^2}
\left[\begin{matrix}1\\0\end{matrix}\right],$$
$$\Psi_{(1;{1\over2},1,{1\over2})}=\left({\kappa\over\pi}\right)^{{3\over4}}\left({2\kappa\over3}\right)^{{1\over2}}
\e^{-{1\over2}\kappa
r^2}\left[\begin{matrix}z\\x+iy\end{matrix}\right],$$
$$\Psi_{(2;{1\over2},0,{1\over2})}=\left({\kappa\over\pi}\right)^{{3\over4}}\left(3\over2\right)^{1\over2}\e^{-{1\over2}\kappa r^2}
\left(1-{2\over3}\kappa
r^2\right)\left[\begin{matrix}1\\0\end{matrix}\right],$$
$$\Psi_{(2;{1\over2},1,{1\over2})}=\left({\kappa\over\pi}\right)^{{3\over4}}\left({5\kappa\over3}\right)^{{1\over2}}
\e^{-{1\over2}\kappa r^2}\left(1-{2\over5}\kappa
r^2\right)\left[\begin{matrix}z\\x+iy\end{matrix}\right],$$ where
$\Psi_{(n;{j},\ell,{m_j})}$ explicitly indicates the important
quantum numbers. Only the first two components of the 4-spinors are
shown. In the subspace spanned by these vectors, $\tilde h^2$ is
represented by the matrix
$$\left[\begin{matrix}3&{2\sqrt{2}\over\sqrt{3\pi}}&0&0\\{2\sqrt{2}\over\sqrt{3\pi}}&5&{2\over3\sqrt{\pi}}&0
\\0&{2\over3\sqrt{\pi}}&7&{7\over3\sqrt{2\over5\pi}}\\0&0&{7\over3
\sqrt{2\over5\pi}}&9\end{matrix}\right]\kappa$$ which
has the following eigenvalues$$9.30521\kappa,~ 6.77892\kappa,~
5.28004\kappa,~ 2.63584\kappa.$$
If we had considered the space spanned by the first two vectors, the
eigenvalues would have been $5.35972\kappa,~ 2.64028\kappa$, showing
a quick convergence. If center of mass corrections are not
considered, the lowest eigenvalue is identified with the quark mass
squared.
$$m_q^2=\left(4-\sqrt{1+{8\over3\pi}}\right)\kappa,\quad \kappa=34087.2
~{\rm MeV}^2,\quad R_M=~
1.38398{\rm fm}.$$

\subsection{Calculation of
$\langle\Phi_0|(\bfp_1+\bfp_2+\bfp_3)^2|\Phi_0\rangle$}
Center of mass corrections must be considered. As previously
observed, the correction for the center of mass (CM) motion is
implemented by subtracting the expectation value of the CM momentum
squared from the expression for the square of the nucleon mass. We
need, therefore,
$\langle\Phi_0|(\bfp_1+\bfp_2+\bfp_3)^2|\Phi_0\rangle$, where
$|\Phi_0\rangle$ is the wave function of the three-quark system.

In the subspace spanned by $\Psi_{(1;{1\over2},0,{1\over2})}$ and
$\Psi_{(1;{1\over2},1,{1\over2})}$, $\tilde h^2$ is represented by
the matrix
$$\left[\begin{matrix}3&{2\sqrt{2}\over\sqrt{3\pi}}\\{2\sqrt{2}\over\sqrt{3\pi}}&5
\end{matrix}\right]\kappa$$ The groundstate eigenvalue reads
$\left(4-\sqrt{1+{8\over3\pi}}\right)\kappa$ and the associated
eigenvector may be written
$\Psi_0=c_1\Psi_{(1;{1\over2},0,{1\over2})}+c_2\Psi_{(1;{1\over2},1,{1\over2})}$.
The mass squared of the nucleon should be identified with
$9\left(4-\sqrt{1+{8\over3\pi}}\right)\kappa$. Taking into account
center of mass correction, means subtracting
$\langle\Phi_0|(\bfp_1+\bfp_2+\bfp_3)^2|\Phi_0\rangle$. We obtain
$$\langle\Psi_{(1;{1\over2},0,{1\over2})}|\bfp^2|\Psi_{(1;{1\over2},0,{1\over2})}\rangle
={3\kappa\over2},\quad
\langle\Psi_{(1;{1\over2},1,{1\over2})}|\bfp^2|\Psi_{(1;{1\over2},1,{1\over2})}\rangle={5\kappa\over2}$$
$$\langle\Psi_{(1;{1\over2},1,{1\over2})}|\bfp|\Psi_{(1;{1\over2},0,{1\over2})}\rangle
=-i\hat k{\sqrt{{\kappa\over6}}},
$$so that {the correct expression reads}
$$\langle\Phi_0|(\bfp_1+\bfp_2+\bfp_3)^2|\Phi_0\rangle
=\left({3\left(3c_1^2+5c_2^2\right)\over
2(c_1^2+c_2^2)}+{c_1^2c_2^2\over
(c_1^2+c_2^2)^2}\right){\kappa}=9\Delta\kappa,$$where
$$c_1={-3\pi-\sqrt{3\pi(8+3\pi)}
},\quad c_2=2\sqrt{6\pi},
$$ the quantity $\Delta\kappa$  being the center of mass correction for each quark,
so that, in this approximation, the quark mass becomes
$m_q^2=\left(\left(4-\sqrt{1+{8\over3\pi}}\right)- Delta \right)\kappa$.
Setting
$m_q=300$ MeV, we find $\kappa=
43197.8$ MeV$^2$.  The nucleon mass radius which is derived from the
expectation value of $\bfr_1^2$ is too big due to the fluctuation of
the nucleon CM.  A reasonable value is obtained if, instead of the
expectation value of $\bfr_1^2$, one compensates for the CM motion
and considers the expectation value of
$(\bfr_1-(\bfr_1+\bfr_2+\bfr_3)/3)^2$. This is equal to the
expectation value of $2( \bfr_1^2- \bfr_2\cdot\bfr_3)/3$. Now,
$${2\over3}\langle\Phi_0|
\bfr_1^2- \bfr_2\cdot\bfr_3|\Phi_0\rangle={2\over3}
\left({3\over2\kappa}c_1^2+{5\over2\kappa}c_2^2+{2\over6\kappa}c_1^2c_2^2\right)$$
so that, for $\kappa=
43197.8$ MeV$^2$ a mass radius equal to $0.930548$ fm is obtained.

\section{QMC model. Bogoliubov model with external scalar field} 
According to the quark-meson coupling (QMC) model \cite{guichon}, 
nuclear matter is
a system of nucleons which behave like point-like particles,
although they are constituted by quarks coupled to the scalar
$\sigma$ field, in the framework of an independent particle model.
The QMC model, which has been proposed by Guichon \cite{guichon} on
the basis of the MIT bag model, has been considered by other authors
\cite{guichon1,whittenbury}. Recently, it has also been implemented
on the basis of a quadratically raising potential
\cite{batista,barik1}. Here, we wish to implement the QMC model
based on the Bogoliubov quark model \cite{bogolubov}.

The energy density of quark matter reads
\begin{eqnarray}&&{\cal E}={\gamma\over(2\pi)^3}\int^{k_F}\d^3k\sqrt{k^2+{M^*}^2}+{1\over2}m_\sigma^2\sigma^2
-{1\over2}m_\omega^2\omega^2+g_\omega\omega\rho_B\label{eq3}\\
&&{\rho_B}={\gamma\over(2\pi)^3}\int^{k_F}\d^3k,\quad
M^*=3{m}_q(\sigma), \nonumber\end{eqnarray} where $\gamma=4$ denotes
the spin isospin degeneracy. Clearly, $\sigma$ and $\omega$ should
be replaced by the values which minimize $\cal E.$

The pressure is given by
$${-P}={\gamma\over(2\pi)^3}\int^{k_F}\d^3k\sqrt{k^2+{M^*}^2}+\rho_B\left(g_\omega\omega-\sqrt{k_F^2+{M^*}^2}\right)+{1\over2}m_\sigma^2\sigma^2
-{1\over2}m_\omega^2\omega^2. 
$$ Clearly,
$\sigma$ and $\omega$ should be replaced by the values which
maximize $P$ for $k_F=\sqrt{\mu^2-{M^*}^2}$. Notice that $M^*$
depends on $\sigma$, but not on $\omega$. {In order to determine
$m_q(\sigma)$, we introduce, following \cite{guichon}, an external
sigma field acting on the quarks, so that, in eq. (\ref{Ha}), the
term $\kappa\sum_{j=1}^3\beta_j|{\bfr}_j| $ is replaced by
$\sum_{j=1}^3\beta_j\left(g_\sigma\sigma+\kappa|{\bfr}_j|\right) $.
Next we determine $m_q(\sigma)=M^*/3$ and the binding energy of
nuclear matter within two approximation schemes.}

\subsection{Operator $\bfp^2+(-a+|\bfr|\kappa)^2$}
It is natural to regard as a perturbation, in eq. (\ref{h^2-N}), the
term involving $\bfalpha\cdot\bfr$, and to restrict our attention to
the operator $\bfp^2+(-a+|\bfr|\kappa)^2$, where $a=g_\sigma\sigma$.
Then,  a simplified model is obtained which may be interesting to
consider because much of the basic physics is, indeed, already in
it.

\begin{figure}[ht]
\vspace{1.5cm} \centering
\includegraphics[width=0.8\linewidth,angle=0]{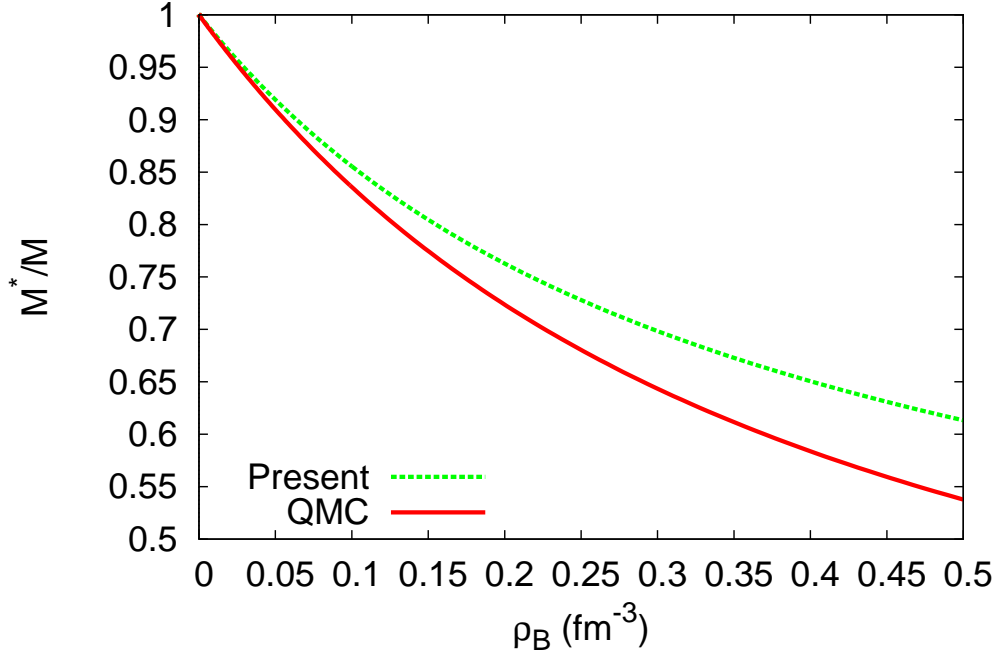}
\caption{Nucleon effective mass for the {present} QMC approach based
on the Bogolyubov model, according to eqs. (\ref{new-full-Delta}).
Comparison with the original QMC model of \cite{guichon}.
} \label{fig8}
\end{figure}
The quark groundstate wave function has
angular momentum $\ell=0$, and, in the presence of an external
scalar field $\sigma$, is given by the ansatz
\begin{equation}\label{l=0}\Psi_{0,0}=\exp\left({-{1\over2}\left(\sqrt{\kappa}r-{a\over\sqrt{\kappa}}\right)^2}\right)
\left[\begin{matrix}1\\0\end{matrix}\right],
\end{equation} 
Let
\begin{eqnarray*}
&&F_{N,0}(\kappa,{a\over\sqrt{\kappa}})=\int\d^3\bfr\Psi^\dag_{0,0}\Psi_{0,0},\\
&&F_{K,0}(\kappa,{a\over\sqrt{\kappa}})=\int\d^3\bfr\Psi^\dag_{0,0}(-\bfnabla^2)\Psi_{0,0},\\
&&F_{P,0}(\kappa,{a\over\sqrt{\kappa}})=\int\d^3\bfr\Psi^\dag_{0,0}(\kappa
r-a)^2\Psi_{0,0}.
\end{eqnarray*}
We  find\begin{eqnarray*}&&F_{N,0}(\kappa,{a\over\sqrt{\kappa}})=\int\d^3\bfr\e^{-\left(\sqrt{\kappa}r-{a\over\sqrt{\kappa}}\right)^2},\\
&&F_{K,0}(\kappa,{a\over\sqrt{\kappa}})=\int\d^3\bfr
\left(3\kappa-{2a\over r}-(\kappa r-a)^2\right)\e^{-{1\over2}
\left(\sqrt{\kappa}r-{a\over\sqrt{\kappa}}\right)^2},\\
&&F_{P,0}(\kappa,{a\over\sqrt{\kappa}})= \int\d^3\bfr (\kappa
r-\alpha)^2\e^{-\left(\sqrt{\kappa}r-{a\over\sqrt{\kappa}}\right)^2.
}\end{eqnarray*}So that
\begin{eqnarray*}&&F_{N,0}(\kappa,\alpha)=
{\pi\over\kappa\sqrt{\kappa}}\left(
2\alpha\e^{-\alpha^2}+(1+2\alpha^2){\sqrt{\pi}}(1+\Erf(\alpha))\right)\\
&&F_{P,0}(\kappa,\alpha)=F_{K,0}(\kappa,\alpha)=
{\pi\over2\sqrt{\kappa}}\left(
2\alpha\e^{-\alpha^2}+(3+2\alpha^2)\sqrt{\pi}(1+\Erf(\alpha))\right).
\end{eqnarray*}where $\alpha={a/\sqrt{\kappa}}={g_\sigma\sigma/\sqrt{\kappa}}$. The expression for
the squared quark mass reads
\begin{equation}m_q^2(\kappa,\alpha)={F_{K,0}(\kappa,\alpha)+F_{P,0}(\kappa,\alpha)\over
F_{N,0}(\kappa,\alpha)},\label{new0}\end{equation} or, if the CM
correction is considered
\begin{eqnarray}m_q^2(\kappa,\alpha)={F_{K,0}(\kappa,\alpha)+F_{P,0}(\kappa,\alpha)\over
F_{N,0}(\kappa,\alpha)}-{F_{P,0}(\kappa,\alpha)\over
3F_{N,0}(\kappa,\alpha)}.\label{new}\end{eqnarray} Minimization of
(\ref{eq3}) with respect to $\sigma$ is easily performed
and the minimizing value of $\sigma$ is determined by requiring
self-consistency.
\begin{figure}[ht]
\vspace{1.5cm} \centering
\includegraphics[width=0.8\linewidth,angle=0]{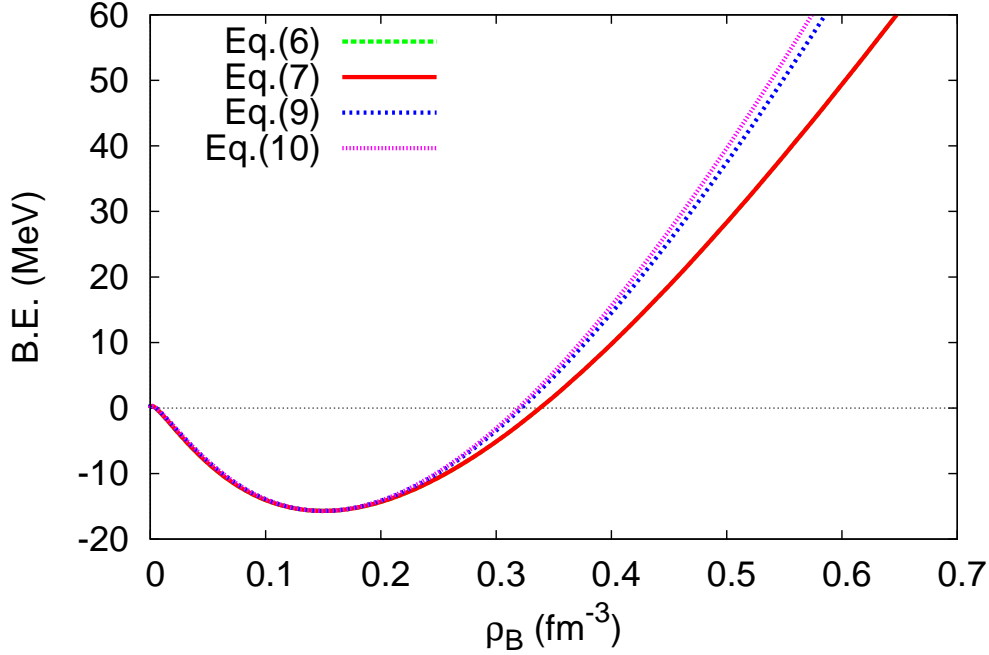}
\caption{Binding energy for the QMC approach based on the Bogolyubov
model, 
{using the effective mass defined by means of} eqs. (\ref{new0}),
(\ref{new}), (\ref{new-full}) and (\ref{new-full-Delta}). The curves
corresponding to eqs. (\ref{new0}) and (\ref{new}) coincide. }
\label{fig9}
\end{figure}

\subsection{Operator $\bfp^2+(-a+|\bfr|\kappa)^2+\sigma_r\kappa$}
{Going beyond the previous Section, we investigate now the effect of
the term involving $\bfalpha\cdot\bfr$ which appears in eq.
(\ref{h^2-N}).} We seek the matrix which represents the operator
$\bfp^2+(-a+|\bfr|\kappa)^2+\sigma_r\kappa$ in a subspace spanned by
$\ell=0$ and $\ell=1$ wave-functions.

In the presence of an external scalar field $\sigma$,
the wave function  of the lowest quark state with angular momentum
$\ell=1$, is given by the ansatz
\begin{equation}\Psi_{0,1}=\exp\left({-{1\over2}\left(\sqrt{\kappa}r-{a\over\sqrt{\kappa}}\right)^2}\right)
\left[\begin{matrix}z\\x+iy\end{matrix}\right].\label{l=1}\end{equation}
In the space spanned by the wave-functions (\ref{l=0}) and
(\ref{l=1}), the operator
$\bfp^2+(-a+|\bfr|\kappa)^2+\sigma_r\kappa$ is represented by the
matrix
$$A=\left[\begin{matrix}a_{00} &a_{01}
\\a_{10}
&a_{11}
\end{matrix}\right],$$
where
\begin{eqnarray*}&&a_{00}={F_{K,0}(\kappa,\alpha)+F_{P,0}(\kappa,\alpha)\over
F_{N,0}(\kappa,\alpha)},\\&&a_{01}=a_{10}={F_{C,01}\over\sqrt{F_{N,0}F_{N,1}}},\\&&a_{11}={F_{K,1}(\kappa,\alpha)+F_{P,1}(\kappa,\alpha)\over
F_{N,1}(\kappa,\alpha)},
\end{eqnarray*}the functions
$F_{K,0},~F_{P,0},~F_{N,0},~F_{K,1},~F_{P,1},~F_{N,1},~F_{C,01},$
being defined and given in the Appendix. The lowest eigenvalue of
$A$ reads
\begin{equation}\label{new-full}m_q^2(\kappa,\alpha)={1\over2}\left(a_{00}+a_{11}-\sqrt{(a_{00}-a_{11})^2+4a_{01}^2}\right).\end{equation}
Here, CM corrections have not been yet included.
\begin{figure}[ht]
\vspace{1.5cm} \centering
\includegraphics[width=0.8\linewidth,angle=0]{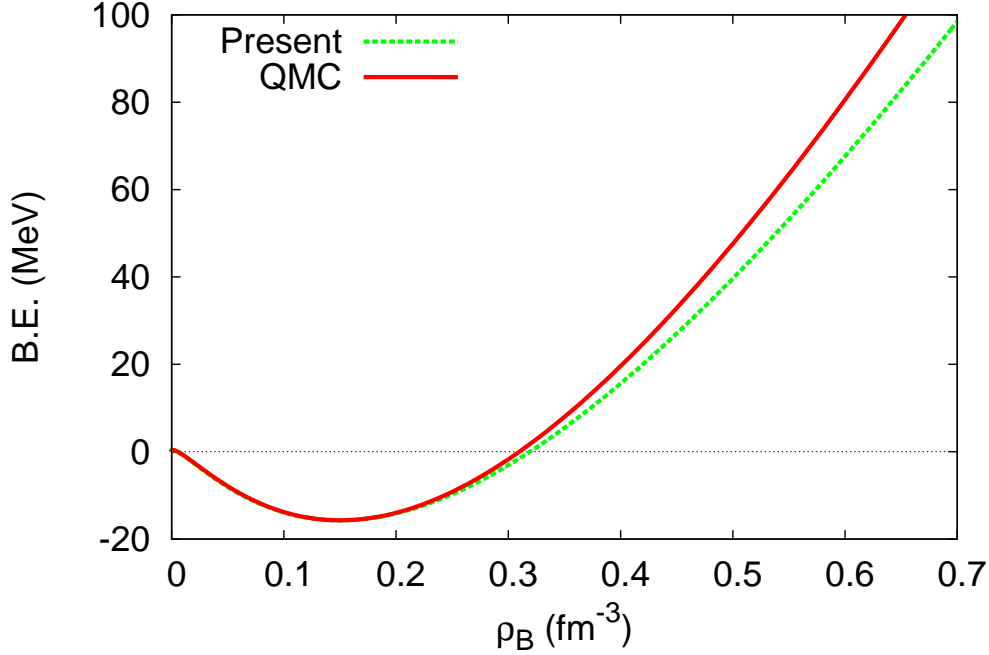}
\caption{Binding energy for the {present} QMC approach based on the
Bogolyubov model, according to eqs. (\ref{new-full-Delta}).
Comparison with the {original QMC} model of \cite{guichon}.
}
\label{fig10}
\end{figure}
In order to take into account the  CM corrections, we need the
integral
\begin{eqnarray*}&&G_{C,01}(\kappa,{a\over\sqrt{\kappa}})={1\over3}\int\d^3\bfr
r(\kappa
r-a)\e^{-(\sqrt{\kappa}r-{a\over\sqrt{\kappa}})^2}\\
&&G_{C,01}(\kappa,{\alpha})={\pi\over2\kappa\sqrt{\kappa}}\left(2\alpha\e^{-\alpha^2}
+(1+2\alpha^2)\sqrt{\pi}(1+\Erf(\alpha))\right).\end{eqnarray*}We
may write
\begin{equation}\label{new-full-Delta}
m_q^2(\kappa,\alpha)={1\over2}\left(a_{00}+a_{11}-\sqrt{(a_{00}-a_{11})^2+4a_{01}^2}\right)-\Delta_{CM},
\end{equation} where
$$\Delta_{CM}={1\over3( c_1^2+c_2^2)}\left({F_{K,0}(\kappa,\alpha)\over
F_{N,0}(\kappa,\alpha)}c_1^2+{F_{K,0}(\kappa,\alpha)\over
F_{N,0}(\kappa,\alpha)}c_2^2\right)+
{6c_1^2c_2^2(G_{C,01}(\kappa,\alpha))^2\over
9(c_1^2+c_2^2)^2F_{N,0}(\kappa,\alpha)F_{N,1}(\kappa,\alpha)},$$with
$$c_1=a_{11}-a_{00}+\sqrt{(a_{11}-a_{00})^2+4a_{01}^2},\quad c_2=-2a_{01}~.$$
Minimization of (\ref{eq3}) with respect to $\sigma$ is easily
performed.
and the minimizing value of $\sigma$ is determined by requiring
self-consistency.
\subsection{Discussion}

In Fig.\ref{fig8}, the nucleon effective mass for the {present} QMC
approach based on
eqs. (\ref{new-full-Delta}) is represented and is compared with the
original QMC model of \cite{guichon}, {showing that the Bogolyubov
model leads to a smoother decrease of $M^*$ with the baryonic
density.} In Fig. \ref{fig9}, the binding energy for the QMC
approach based on the Bogolyubov model, {using the effective mass
defined my means of} eqs. (\ref{new0}), (\ref{new}),
(\ref{new-full}) and (\ref{new-full-Delta}) is shown. { We observe
that the curves corresponding to eqs. (\ref{new0}) and (\ref{new})
coincide. The effect of the $\kappa\sigma_r$ term is to stiffen the
EOS, as expected.} The curve displayed in Figure 3, representing the
binding energy vs. the baryon density, was obtained using
(\ref{new-full-Delta}) {and shows that the Bogolubov model leads to
a slightly less stiff EOS than the original QMC model of
\cite{guichon}.} The inputs are $m_\sigma=$550 MeV, $m_\omega=783$
MeV, $m_q=313$ MeV. The coupling constants $g_\sigma,~g_\omega$
  were chosen so as to reproduce the binding energy and density at
equilibrium, that is, ${\cal E}/\rho_B - M_N = -15.7$~
  MeV at saturation    (pressure P=0), being
$M_N=939$ MeV the free nucleon mass. The value of $\kappa$  is
determined by the quark mass  in vacuum, $m_q$. The effective quark
mass at saturation density is denoted by $m_q^*$. The outputs are
summarized in Table 1.

\section{Comparison with the Walecka and Zimanyi-Moszkowski models}

It is challenging to compare  the nucleon effective mass, as
predicted by the present version of the QMC model, and  by the
Walecka and the Zimanyi-Moszkowski models 
\cite{CW,ZM}, with respect to the dependence on the scalar field. We
consider the expression of the effective mass in terms of the scalar
coupling $g_\sigma\sigma.$ In the Walecka model, the relation
$M_{eff}=M(\sigma)=1-g_\sigma\sigma$ holds. The mass of the free
nucleon is set equal to 1. In the derivative coupling model (cf.
\cite{ZM}), we find
$$M_{eff}={1/(1+g_\sigma\sigma})=1-g_\sigma\sigma+(g_\sigma\sigma)^2-(g_\sigma\sigma)^3+\ldots.$$
In the simple version of the QMC model obtained in the present
formulation, eq. (\ref{new0}), we have, to a good approximation, {a
close expression to the previous one, if cubic and higher terms in
$g_\sigma\sigma$  are neglected.} According to both the present
version {of QMC} and the bag version of QMC ({cf.\cite{guichon}})
these terms are very small. In particular, in the present version,
it turns out that the coefficient of the squared term in an
expansion of the effective mass is an order of magnitude less than
it is in the ZM model. Specifically, {keeping in mind that in the
groundstate the expectation values of $\bfp^2$ and
$\kappa(|\bfr|-a)^2$ are equal, and taking $\kappa=b^{-2}$}, we find
that:
\begin{eqnarray*}&&M_{eff}=\sqrt{\displaystyle{2\int_0^\infty
r^2(r-a)^2\exp(-(r-a)^2b^{-2})\d
r}\over\displaystyle{3b^2\int_0^\infty r^2\exp(-(r-a)^2b^{-2})\d
r}}\\&&=1-0.3761\alpha+0.1113\alpha^2-0.00294\alpha^3+\ldots\end{eqnarray*}
where $\alpha=a/b$. We have for this case,
$$g_\sigma\sigma={2\alpha\over3\sqrt{\pi}}\approx0.3761\alpha,\quad M_{eff}
=1-g_\sigma\sigma+0.788(g_\sigma\sigma)^2-0.131(g_\sigma\sigma)^3+\ldots.$$
{which is close to
$M_{eff}={1/(1+g_\sigma\sigma)}=1-g_\sigma\sigma+(g_\sigma\sigma)^2.$}
For large number of dimensions $D \gg 1$, {that is, if in the
previous integrations over $r$, $r\d r$ is replaced by $r^{D-1}\d
r$}, we obtain:
$$M_{eff}=1-g_\sigma\sigma+{1\over2}(g_\sigma\sigma)^2,$$
where
$$g_\sigma\sigma={1\over\sqrt{2D}}\alpha.$$
Still, it is most remarkable that the present expression is very
close to the corresponding Zimanyi-Moszkowski expression. Actually,
both expressions are in agreement up to the mentioned cubic order.
The difference shows up in the ratio $K/|W_0|$ , between the
incompressibility and the binding energy  $|W_0|=|{\cal E}/\rho_B -
M_N|$. In the weak coupling limit, we have in lowest order,
$K/|W_0|=18(1-2\sqrt{|W_0|/(M_Nc^2)})$ for {the present version of
QMC}, and $K/|W_0|=18(1-3\sqrt{|W_0|/(M_Nc^2)})$ for
Zimanyi-Moszkowski model. Thus for the same binding energy, {the
present version of QMC} leads to a slightly larger $K$ than
Zimanyi-Moszkowski model. The incompressibility of 235 MeV which is
found, is close to the value of 225 MeV of the derivative coupling
model, and the effective mass is 0.85 for both models. For the more
realistic version, eq. 
(\ref{new-full-Delta}), things are very close.
\begin{table*}[t]
\centering
\renewcommand{\arraystretch}{1.4}
\setlength\tabcolsep{3pt}
\begin{tabular}{|c|c|c|c|c|}
\hline
&  eq (6) & eq (7)& eq (9) & eq (10)\\
\hline
$g_\sigma$& 3.876 & 4.246 & 3.696 &4.024\\
$g_\omega$& 6.492 & 6.492 & 7.818 & 8.159\\
$\sigma$  (MeV)& 20.57 & 20.57 & 23.76& 24.60\\
$\omega$  (MeV)& 12.20 & 12.20 &14.69&15.33\\
$M_N^*$ (MeV) & 803.84 & 803.84& 766.30& 755.49\\
K (MeV) & 235.5 & 235.5 & 245.9
&249.1\\
$\kappa$
(MeV$^2)$&$3.27\times10^4$&$3.92\times10^4$&$3.71\times10^4$&$4.60\times10^4$\\
$m_q^*/ m_q$&0.95&0.85&0.81&0.80 \\\hline
\end{tabular}
\caption{\label{table1} Outputs for nuclear matter, which have been
determined from the binding energy at equilibrium density, being
$M_N^*$ the corresponding effective nucleon mass.}
\end{table*}
\section{Conclusions}

In the present work we have proposed an effective relativistic
nuclear model that takes into account the internal structure of the
nucleon explicitly, in the philosophy of the QMC model of Guichon
\cite{guichon}. Matter at low densities and temperatures is a system
of nucleons composed of quarks bound by a linearly raising
potential, as suggested by gauge theories, according to the
Bogoliubov  model of baryons \cite{bogolubov}. The parameters of the
model have been fitted to the saturation density and the binding
energy of symmetric nuclear matter at this density, and the quark
mass in vacuum. As output the incompressibility of matter  and the
effective nucleon mass at saturation were calculated respectively
with values 249.1  MeV and 0.8 $M$.

The string tension which is obtained with Bogoliubov model turns out
to be less than one fourth  of the string tension which is found in
the lattice calculation \cite{bissey}. However, this model is
incomplete since it does not account for the mass splitting between
the nucleon and the   resonance. If the model is refined in this
direction, following, for instance, \cite{barik,chen}, a higher
string tension, closer to $0.5$ GeV fm$^{-1}$ may be obtained. Such
improvement is left for a future publication.

On the other hand, it may be observed that this model is suggested
by the string concept and leads essentially to a 3-dimensional
harmonic oscillator, since the operator $h^2$ may be approximated by
$\bfp^2+\kappa^2\bfr^2$. The failure of the model in accounting for
the correct string tension is, perhaps, the price one has to pay for
the approximate treatment of the string force. Indeed, the
Bogoliubov model is based on the idea that the quarks move
independently in a three dimensional potential. Now, a string is a
one-dimensional object. If we replace the 3-dimensional oscillator
by a 1-dimensional oscillator, the value of the string tension which
reproduces the quark mass increases. Indeed, consider the
1-dimensional harmonic oscillator $p_z^2+\kappa^2z^2$ . The lowest
eigenvalue of this operator,  which must be identified with the
quark mass squared, is $\kappa$. Since the average nucleon$-$Delta
resonance mass is (1232 + 939)/2 MeV =1085.5 MeV, the quark mass is
$\sqrt{\kappa}=361.8$ MeV$^2$=$0.655$ GeV/fm, so that
  0.655 GeV/fm, which is a sizable improvement over the simple 3D model.

\section*{Appendix}
Let
\begin{eqnarray*}
&&F_{N,1}(\kappa,{a\over\sqrt{\kappa}})=\int\d^3\bfr\Psi^\dag_{0,1}\Psi_{0,1},\\
&&F_{K,1}(\kappa,{a\over\sqrt{\kappa}})=\int\d^3\bfr\Psi^\dag_{0,1}(-\bfnabla^2)\Psi_{0,1},\\
&&F_{P,1}(\kappa,{a\over\sqrt{\kappa}})=\int\d^3\bfr\Psi^\dag_{0,1}(\kappa
r-a)^2\Psi_{0,1},\\
&&F_{C,01}(\kappa,{a\over\sqrt{\kappa}})=\int\d^3\bfr
\Psi^\dag_{0,0} {\kappa\over
r}\left[\begin{matrix}z&x-iy\\x+iy&-z\end{matrix}\right]\Psi_{0,1}.
\end{eqnarray*} We find
\begin{eqnarray*}
&&F_{N,1}(\kappa,{a\over\sqrt{\kappa}})=\int\d^3\bfr r^2\e^{-\left(\sqrt{\kappa}r-{a\over\sqrt{\kappa}}\right)^2},\\
&&F_{K,1}(\kappa,{a\over\sqrt{\kappa}})=\int\d^3\bfr r^2
\left(5\kappa-{4a\over r}-(\kappa r-a)^2\right)\e^{-{1\over2}
\left(\sqrt{\kappa}r-{a\over\sqrt{\kappa}}\right)^2},\\
&&F_{P,1}(\kappa,{a\over\sqrt{\kappa}})= \int\d^3\bfr r^2(\kappa
r-\alpha)^2\e^{-\left(\sqrt{\kappa}r-{a\over\sqrt{\kappa}}\right)^2}\\
&&F_{C,01}(\kappa,{a\over\sqrt{\kappa}})=\kappa\int\d^3\bfr
r\e^{-\left(\sqrt{\kappa}r-{a\over\sqrt{\kappa}}\right)^2},
\end{eqnarray*} so that,
\begin{eqnarray*}&&
F_{N,1}(\kappa,{\alpha})
={\pi\over2\kappa^{(5/2)}}\left((10\alpha+4\alpha^3)\e^{-\alpha^2}
+(3+12\alpha^2+4\alpha^4)\sqrt{\pi}(1+\Erf(\alpha))\right)\\
&&F_{K,1}(\kappa,{\alpha}) = F_{P,1}(\kappa,{\alpha})
={\pi\over{4}\kappa^{3/2}}\left((34\alpha+4\alpha^3)\e^{-\alpha^2}
+(15+36\alpha^2+4\alpha^4)\sqrt{\pi}(1+\Erf(\alpha))\right)\\
&&F_{C,01}(\kappa,{\alpha})= {\pi\over\kappa}
\left(2(1+\alpha^2)\e^{-\alpha^2}+(3\alpha+2\alpha^3){\sqrt{\pi}}(1+\Erf(\alpha))\right)
\end{eqnarray*}
\end{document}